# Divergent effects of static disorder and hole doping in geometrically frustrated $\beta$-CaCr$_2$O$_4$

S. E. Dutton[1*], C. L. Broholm[2], and R. J. Cava[1]

[1]Department of Chemistry, Princeton University, Princeton, NJ 08544

[2]Department of Physics and Astronomy, The Johns Hopkins University, Baltimore, MD 21218

**Abstract**

The gallium substituted and calcium deficient variants of geometrically frustrated $\beta$-CaCr$_2$O$_4$, $\beta$-CaCr$_{2-2x}$Ga$_{2x}$O$_4$ (0.02$\leq$ x$\leq$ 0.25) and $\beta$-Ca$_{1-y}$Cr$_2$O$_4$ (0.075$\leq$ y$\leq$ 0.15), have been investigated by x-ray powder diffraction, magnetization and specific heat measurements. This allows for a direct comparison of the effects, in a geometrically frustrated magnet, of the static disorder that arises from non-magnetic substitution and the dynamic disorder that arises from hole doping. In both cases, disturbing the Cr$^{3+}$ lattice results in a reduction in the degree of magnetic frustration. On substitution of Ga, which introduces disorder without creating holes, a gradual release of spins from ordered antiferromagnetic states is observed. In contrast, in the calcium deficient compounds the introduction of holes induces static ferrimagnetic ordering and much stronger perturbations of the $\beta$-CaCr$_2$O$_4$ host.

* Corresponding Author. Tel: +1609 258 5556 Fax: +1609 258 6746. E-mail address: siandutton@cantab.net

## 1. Introduction

Competing interactions and low dimensionality can suppress conventional magnetic order in insulating materials[1-9], resulting in the absence of static long-range order below the Weiss temperature, $\theta$. In these strongly correlated regimes, when the magnetic interactions exceed the thermal energy scale (T< $\theta$), weak symmetry breaking interactions and structural disorder can play an essential role. While the effects of dilution on the magnetic lattice by non-magnetic cations have been studied[5,9] much less is known about the effects of hole or electron doping. Here we report the surprisingly strong effects of hole doping the quasi-one-dimensional geometrically frustrated magnet β-$CaCr_2O_4$. An increase in the electrical conductivity indicates that the holes can hop from site to site. We find the impact of such potentially dynamic bond centred disorder is far greater than that associated with non-magnetic disruption of the same magnetic lattice.

The magnetic properties of β-$CaCr_2O_4$[10] were recently reported[2]. This compound adopts the $CaFe_2O_4$ structure type[11], in which two leg ladders of edge sharing $CrO_6$ octahedra (creating chains of edge sharing magnetic triangles) propagate along the $c$ axis. The presence of two inequivalent Cr sites results in two distinct but similar chains (shown in green and blue in Figure 1). The two triangle-based magnetic chains share corners with each other, generating honeycomb tunnels occupied by Ca. In the resultant $Cr^{3+}$ ($d^3$) magnetic lattice, multiple competing interactions are present[2]. On cooling below 270 K, β-$CaCr_2O_4$ first displays an extended quasi-one-dimensional magnetic regime followed below 21 K by three-dimensional magnetic ordering[2]. A change in the 3D ordering occurs below 17 K in which long-wavelength elliptical cycloid spin modulations propagate along the length of the ladders (propagation vector **k**= (0, 0, 0.477)), with the magnetic moments aligned within the $ac$-plane. This non-trivial ordering scheme is a consequence of the magnetic frustration, specifically competing nearest and next nearest neighbour exchange interactions along the chains. It also appears that Dzyaloshinskii-Moriya interactions stabilize definite spin chirality on each zig-zag spin ladder. Given the low dimensional and frustrated nature of magnetism in β-$CaCr_2O_4$, we expect that small changes in the nature of the magnetic framework may have a significant effect on the low temperature ordering behaviour. Manipulating these interactions through chemical substitution is the goal of the current study. The existence of a high pressure phase of $CaGa_2O_4$[12, 13] with the $CaFe_2O_4$ structure suggests that dilution of the magnetic lattice with non-magnetic $Ga^{3+}$ should be possible. During the course of this study we found that materials prepared in non-reducing conditions are black in color (stoichiometric β-$CaCr_2O_4$ is bright green). This may indicate either the presence of $Cr^{4+}$, i.e. hole doping of the Cr sublattice, or, alternatively, holes on the oxygen sublattice, i.e. $O^-$ on the superexchange mediating oxygen sites Here we present a comparison of the properties of β-$CaCr_{2-2x}Ga_{2x}O_4$ and β-$Ca_{1-y}Cr_2O_4$ with, where possible, equivalent concentrations of $Cr^{3+}$. This allows us to distinguish the effects of hole doping from the effects of disorder in the magnetic lattice. We find that the effects are surprisingly divergent.

## 2. Experimental

Powder samples of β-$CaCr_{2-2x}Ga_{2x}O_4$ (0.01≤ x≤ 0.25) and β-$Ca_{1-y}Cr_2O_4$ (0.075≤ y≤ 0.150) were prepared using a ceramic synthesis route. Stoichiometric quantities of $CaCO_3$ (Alfa Aesar 99.99%), $Cr_2O_3$ (Alfa Aesar 99.97%) and pre-dried $Ga_2O_3$ (Alfa Aesar 99.99%) were intimately mixed and fired at 1000 °C in air as a powder overnight. Pellets were subsequently heated repeatedly at 1200 °C in air for 24 hours. For the gallium free samples this process yielded single phase, black samples of β-$Ca_{1-x}Cr_2O_4$ for 0.075≤ x≤ 0.15. Single phase samples of β-$CaCr_{2-2x}Ga_{2x}O_4$ (0.01≤ x≤ 0.25) were obtained by further heating for 12 hours under Ar in a vacuum furnace at 1200 °C. Undoped β-$CaCr_2O_4$ was prepared using a vacuum furnace. Stoichiometric amounts of $CaCO_3$ (Alfa Aesar 99.99%) and $Cr_2O_3$ (Alfa Aesar 99.97%) were intimately mixed. The pellet was slowly heated at 50 °C h$^{-1}$ to 1000 °C under a dynamic vacuum of <10$^{-5}$ Torr. After dwelling at 1000 °C for 2 hours, the furnace was turned off and once at room temperature the chamber filled with Ar. On heating to 1550 °C for 12 hours, phase pure β-$CaCr_2O_4$ was formed.

X-ray powder diffraction patterns were collected on a Bruker D8 Focus operating with Cu Kα radiation and a graphite diffracted beam monochromator. Data were collected over the angular range 5 ≤ 2θ≤ 60 ° with a step size of Δ2θ=0.04 °. Rietveld refinement[14] of the structures was carried out using the GSAS suite of programs[15]. Backgrounds were fitted using a Chebyshev polynomial of the first kind and the peak shape was modeled using a pseudo-Voigt function.

Magnetic susceptibility, specific heat and electrical conductivity measurements were made using a Quantum Design Physical Properties Measurement System (PPMS). Temperature dependent magnetization measurements for finely ground samples were collected under both 0.1 T and 1 T fields between 1.8 and 300 K after cooling in zero field. Isothermal magnetization measurements were performed after cooling in zero field at 5 K, 15 K and 40 K for β-$Ca_{0.925}Cr_2O_4$, spanning the magnetic field range -9 ≤ $μ_0H$ ≤ 9 T, additional data over a narrower field range, 0≤ $μ_0H$ ≤ 2 T, were also collected at higher temperatures. Isothermal magnetization measurements on the β-$CaCr_{2-2x}Ga_{2x}O_4$ (0.01≤ x≤ 0.25) samples spanning the magnetic field range 0≤ $μ_0H$ ≤ 9 T were carried out at 2 K after cooling in zero field. The specific heat of selected samples was measured in zero field over a narrow temperature range spanning the region of interest. Electrical conductivity measurements of β-$Ca_{0.85}Cr_2O_4$ were made in zero field over the temperature range 2≤ T ≤ 300 K using a standard four-probe configuration with contacts made using silver paste. The resistivity of the electrical insulators β-$CaCr_{2-2x}Ga_{2x}O_4$ (0.01≤ x ≤ 0.25) and the parent compound β-$CaCr_2O_4$ were beyond the range of our apparatus (ρ>2 MΩ cm). Pellets for heat capacity and electrical conductivity were prepared by heating for 12 hours at 1250 °C either in air or under Ar in a vacuum furnace for the β-$Ca_{1-y}Cr_2O_4$ and the β-$CaCr_{2-2x}Ga_{2x}O_4$ compositions respectively.

## 3. Results and Discussion

### 3.1 β-$CaCr_{2-2x}Ga_{2x}O_4$

Phase pure samples of β-$CaCr_{2-2x}Ga_{2x}O_4$ were successfully synthesized with up to 25% substitution of Ga on the Cr site; samples with x>0.25 were multiple phase. X-ray diffraction data was analysed in the orthorhombic *Pbnm* space group and the lattice parameters obtained from Rietveld analysis are detailed in Table 1 and Figure 2. The changes in lattice parameters in the β-$CaCr_{2-2x}Ga_{2x}O_4$ series are relatively small. However, the *a* parameter decreases significantly as gallium is added. Since there is only a slight increase in the *b* parameter, the projection of the *ab*-plane, which reflects the shape of the honeycomb lattice, also decreases with increasing x. Conversely, *c* increases slightly with Ga substitution, such that the volumes for x = 0 (285.75(2) Å$^3$) and x = 0.25 (286.04(7) Å$^3$) are almost within error of one another. These small changes reflect the structural similarities between the end members of the solid solution, β-CaCr$_2$O$_4$ and CaGa$_2$O$_4$, particularly the size of the CrO$_6$[2] (<Cr-O>~2.01 Å) and GaO$_6$[13] (<Ga-O>~2.01 Å) octahedra, and are consistent with the absence of significant changes in the lattice parameter in the ZnCr$_x$Ga$_{2-x}$O$_4$ spinel solid solution[16].

Magnetic susceptibility ($\chi=\delta M/\delta H$) for the β-CaCr$_{2-2x}$Ga$_{2x}$O$_4$ series shows no field dependence up to 1 T; hence the data collected in a 1 T field are representative and presented in Figure 3. Fits to the Curie-Weiss law, $\chi=C/(T-\theta)$, were carried out for all compositions for T>200 K. The magnetic parameters obtained are detailed in Table 1. The effective moment per chromium ($\mu_{eff} \sim \sqrt{8C}$) decreases slightly, and the Weiss temperature ($|\theta|$) decreases more significantly with increasing x in β-CaCr$_{2-2x}$Ga$_{2x}$O$_4$ (Figure 4). Deviations from Curie-Weiss behavior below 280 K in β-CaCr$_2$O$_4$[2] have been attributed to low dimensional antiferromagnetic short-range order. Thus fitting to the Curie-Weiss law may not be strictly valid in the temperature regime considered. For x>0.15 the magnetic moment obtained is similar to that expected for spin only Cr$^{3+}$ (3.87 $\mu_B$) suggesting that for these compositions fitting to the Curie-Weiss law is valid. The decrease in effective moment and the magnitude of the Weiss temperature with Ga substitution are consistent with suppression of this quasi-one-dimensional short-range order in β-CaCr$_{2-2x}$Ga$_{2x}$O$_4$.

On substitution of Ga for Cr in β-CaCr$_{2-2x}$Ga$_{2x}$O$_4$, the temperature dependent susceptibility (Figure 3) and specific heat (Figure 5) change systematically. The magnetic susceptibility increases with increasing Ga concentration, such that for β-CaCr$_{1.5}$Ga$_{0.5}$O$_4$ the susceptibility when T<10 K is more than three times that of β-CaCr$_2$O$_4$. At high temperatures, a crossover from predominantly antiferromagnetic to predominantly ferromagnetic correlations is observed in the susceptibility of all the β-CaCr$_{2-2x}$Ga$_{2x}$O$_4$ compositions. This temperature, $T_{co}$, defined as the maximum of $d\chi^{-1}/dT$, increases from ~25 K in β-CaCr$_2$O$_4$ to ~67 K in β-CaCr$_{1.5}$Ga$_{0.5}$O$_4$ (Figure 4). When 0.00≤ x≤ 0.025 a broad maximum in the temperature dependent magnetic susceptibility is observed at ~80 K; with increasing Ga concentration this maximum in the susceptibility becomes less well defined, though it is still present at x=0.050. A large increase in the susceptibility and a low temperature feature, $T_N$, observed as a maximum in $d\chi/dT$, are observed at higher x. As more Ga is introduced and the overall magnetic connectivity decreases, $T_N$ decreases from ~16.5 K for x =0.050 to ~9 K for x =0.25. In the specific heat, two transitions are observed at low values of x. With increasing Ga concentration, the temperatures at which the two transitions in the specific heat occur converge and the transitions themselves become less well defined. Below 10 K an increase in the magnetic entropy is observed in the Ga-rich compositions (x≥ 0.050).

Since the changes to the dimensions of the β-CaCr$_2$O$_4$ structure on Ga substitution are minimal, the nature of the Cr-O-Cr and Cr-Cr interactions that govern the magnetic behavior should not change substantially across the series. The changes in the magnetism in β-CaCr$_{2-2x}$Ga$_{2x}$O$_4$ must therefore primarily be a result of magnetically diluting the Cr$^{3+}$ network. At low temperatures the magnetic susceptibility in the β-CaCr$_{2-2x}$Ga$_{2x}$O$_4$ compositions is larger than for the undoped material. This reflects a suppression of antiferromagnetic ordering due to the disorder introduced by Ga substitution and the release of spins from an antiferromagnetically correlated state. The reversibility of the isothermal magnetisation measured at 2 K (Figure 3 inset) shows that unlike in the gallium doped zinc chromate spinel system[17] the introduction of non-magnetic cations into the β-CaCr$_2$O$_4$ lattice does not result in spin-glass behaviour. The increase in the magnetic susceptibility as Ga is substituted for Cr is thus a result of uncompensated Cr$^{3+}$ cations. The magnetic properties of samples with x<0.050 change slightly from the undoped material. For x≥ 0.050, the observation of a transition at $T_N$ in the susceptibility is a significant deviation from β-CaCr$_2$O$_4$. An indication of the nature of this anomaly is given by the behaviour of the intermediate x=0.050 composition, where the transition in the magnetic susceptibility occurs at the same temperature as the lower temperature peak in the specific heat. This suggests that the lower temperature magnetically ordered structure in Ga doped β-CaCr$_2$O$_4$ is similar to that reported for the undoped composition. Low temperature neutron diffraction would be required to confirm this hypothesis.

### 3.2 β-Ca$_{1-y}$Cr$_2$O$_4$

The range of calcium deficient compositions β-Ca$_{1-y}$Cr$_2$O$_4$ was significantly more restricted than for the Ga substituted series. Using the synthetic method employed, the Ca deficiency for y<0.075 could not be controlled and for y>0.150, a Cr$_2$O$_3$ impurity phase was detected. At the limiting β-Ca$_{0.85}$Cr$_2$O$_4$ composition, 15% of the calcium sites in the tunnels are vacant; these vacancies appear to reduce the phase stability. Heating single phase β-$Ca_{1-y}Cr_2O_4$ (0.075≤ y≤ 0.150) samples under Ar resulted in the formation of stoichiometric β-$CaCr_2O_4$ plus $Cr_2O_3$, indicating that oxidising conditions allowing the presence of holes are critical for the stability of the calcium deficient phase. As in the Ga-substituted series, the x-ray diffraction data were analysed in the orthorhombic *Pbnm* space group and the lattice parameters obtained from Rietveld analysis are detailed in Table 2 and Figure 2. The calcium deficient series shows a more pronounced change in dimension than the Ga series, with all three lattice parameters decreasing as a function of y. This trend is to be expected given the smaller size of $Cr^{4+}$ cations (<Cr-O>=1.91 Å in *Pbnm* $CaCrO_3$[18]) compared to $Cr^{3+}$ (<Cr-O>=1.97 Å in *Pbnm* $LaCrO_3$[19]). The resistivity of β-$Ca_{0.85}Cr_2O_4$ at room temperature is four orders of magnitude lower than that of the insulating undoped and Ga-doped materials. The temperature dependent resistivity follows the activated behaviour expected for a doped semiconductor, with a band gap of 0.57 eV (Figure 6). The positive Seebeck coefficient (~200 μV $K^{-1}$) at room temperature is indicative of hole-type conductivity.

The magnetic susceptibility of the calcium deficient β-$Ca_{1-y}Cr_2O_4$ compositions (Figure 7) is significantly different from that of stoichiometric β-$CaCr_2O_4$. At high temperatures, T>200 K, M vs. H is linear to fields in excess of 1 T and fits to the Curie-Weiss law were carried out on the inverse magnetic susceptibility measured in a 1 T field and are detailed in Table 2. Both the effective magnetic moment and the Weiss temperature are smaller in the Ca-deficient phases than in undoped β-$CaCr_2O_4$. The decrease in the effective magnetic moment is significant, from 4.12 $μ_B$ $mol_{Cr}^{-1}$ in β-$CaCr_2O_4$ to 3.29 $μ_B$ $mol_{Cr}^{-1}$ in β-$Ca_{0.9}Cr_2O_4$ and is less than the expected spin only value for a mixed valence $Cr^{3+}/Cr^{4+}$ composition. This may be due to the persistence of short-range ordering at high temperatures, rendering fits to the Curie-Weiss law inaccurate in the temperature range of our measurements. The independence of the effective magnetic moment to the degree of calcium deficiency is consistent with both the presence of short-range ordering at higher temperatures or holes residing predominantly on $O^-$ sites.

As the Ca deficiency increases, the magnitude of the susceptibility (measured at $μ_0H$=0.1 T in Figure 7) increases; this is unexpected in a local picture for an unfrustrated system given the reduced moment of $Cr^{4+}$ ($d^2$, S=1) relative to $Cr^{3+}$ ($d^3$, S=3/2). The increased susceptibility is consistent with increasing ferromagnetic and/or Dzyaloshinskii-Moriya interactions in the hole doped compositions. The decrease in the frustration index, $f=|\theta|/T_N$, to $f\sim 6.5$ in $\beta$-$Ca_{0.85}Cr_2O_4$ is also consistent with such interactions, which tip the balance from frustrated magnetic interactions towards a potentially heterogeneous ferrimagnetic or canted spin state. While the Cr-Cr distances within both the rungs and the spines decrease with increasing Ca deficiency, there are, however, no significant perturbations to the Cr-O-Cr bond angles that govern the superexchange interactions. Thus changes in the magnetic interactions are likely a result of the introduction of holes into the frustrated magnetic lattice rather than a consequence of structural perturbations. If the holes are introduced into the Cr sites this could result in ferromagnetic clusters within the antiferromagnetically interacting spins. Alternatively holes on the oxygen sub-lattice would change the signs of the superexchange interactions as has been observed in $Y_2BaNiO_5$[20]. Furthermore the calcium vacancies may induce Dzyaloshinskii-Moriya interactions due to locally broken inversion symmetry, leading to a canted antiferromagnetic structure with a ferromagnetic component.

In all hole doped compositions, a maximum in $d\chi/dT$ is observed at ~4 K. At higher temperatures, two further maxima in $d\chi/dT$ are observed; the first decreases in temperature as the calcium deficiency increases, from 19 K in $\beta$-$Ca_{0.925}Cr_2O_4$ to 12 K in $\beta$-$Ca_{0.85}Cr_2O_4$, and the second is observed at 36 K for x=0.075 and x=0.15, and at 30 K in $\beta$-$Ca_{0.9}Cr_2O_4$ (Figure 7). As for the Ga doped samples, a crossover temperature, $T_{co}$, at ~67 K is observed in all $\beta$-$Ca_{1-y}Cr_2O_4$ compositions. Specific heat measurements of $\beta$-$Ca_{0.925}Cr_2O_4$ show a transition at ~18 K (Figure 5) coincident with a maxima in $d\chi/dT$. At 4 K and 36 K, where maxima in $d\chi/dT$ are also noted for $\beta$-$Ca_{0.925}Cr_2O_4$, no features are observed in the specific heat and thus the origin of these anomalies is unclear. At T < 12 K, the magnetic entropy of $\beta$-$Ca_{0.925}Cr_2O_4$ is similar to that of $\beta$-$CaCr_{1.98}Ga_{0.02}O_4$, which in turn has a magnetic entropy similar to that of the parent state. By analogy to $\beta$-$CaCr_2O_4$, the transition at 18 K may involve long-range magnetic ordering and the higher temperature anomaly at ~36 K may be associated with short-range ordering. Isothermal magnetization measurements of $\beta$-$Ca_{0.925}Cr_2O_4$ at 40 K, 60 K and 80 K (Figure 8), i.e. above the higher temperature ordering transition at ~36 K, all show non-linear behavior - only at 100 K is the linear behavior ($\mu_0H<2$ T) expected for paramagnetic behavior observed. From an Arrott plot[21] of the higher temperature data (Figure 8 inset) the ferrimagnetic transition appears to occur at ~70 K, consistent with a diffuse magnetic transition occurring at $T_{co}$. As in the data collected at 40 K, non-linear behavior is observed in the isothermal magnetization measurements collected at 5 and 15 K (Figure 8). The shape of the two lower temperature measurements are similar; in the 5 K measurement a small amount of hysteresis at low fields, $\mu_0H<2$ T, is observed. However compared to the 40 K data the ferrimagnetic transitions observed at 5 and 15 K are much broader, indicating some suppression of the ferromagnetic correlations on cooling below 40 K. The behavior of the hole doped compositions appears to be complex at T<100 K, and further experiments are required to elaborate the various magnetic regimes formed.

**3.3 Comparison of static disorder and hole doping**

Direct comparison of the magnetic susceptibility of samples of β-CaCr$_{2-2x}$Ga$_{2x}$O$_4$ and β-Ca$_{1-y}$Cr$_2$O$_4$ containing equivalent quantities of Cr$^{3+}$ on the magnetic lattice (i.e. x=y) reveals significant differences. An example is shown in Figure 9 for compositions containing 92.5% Cr$^{3+}$. The magnetization of the calcium doped samples is such that when plotted on the same scale the low temperature features in the β-CaCr$_{2-2x}$Ga$_{2x}$O$_4$ compositions can no longer be clearly seen. However, at T>250 K the magnetic susceptibilities converge.

A more informative method for comparing the two series is to rearrange the Curie-Weiss law, to give a dimensionless plot, $\frac{C}{(|\theta|\chi)} = \frac{T}{|\theta|} + 1$, which allows samples with significant differences in their magnetic susceptibility to be compared directly[22]. In this presentation of the data, perfect Curie-Weiss behavior results in a slope of 1 with a y-intercept of 1 for θ<0 (-1 for θ>0); positive deviations from Curie Weiss behavior indicate enhanced antiferromagnetic type fluctuations or short-range order and negative deviations ferromagnetic fluctuations or short-range order. The normalized temperature, T/|θ|, plotted on the x-axis, is scaled relative to the magnitude of the internal magnetic interactions (and is the inverse of the frustration index, f=|θ|/T$_N$).

In the normalized plots (Figure 10) the differences between the two types of doping are highlighted. The deviation from the Curie-Weiss law in the Ca deficient samples occurs at a significantly higher relative temperature, $T/|\theta|>1.5$, compared to the Ga doped samples, where the deviations occur at $T/|\theta|<1$. In the Ga doped samples, relatively small antiferromagnetic deviations from the Curie-Weiss law are observed, whereas in the hole doped compositions, large ferromagnetic deviations are seen. In $\beta$-$Ca_{1-y}Cr_2O_4$, two distinct behaviors in the normalized inverse susceptibility are observed; for y=0.075 the behavior deviates from the Curie-Weiss law for $T/|\theta| < 1.5$, whereas in the more Ca deficient samples this deviation occurs at higher relative temperature, $T/|\theta| < 2.5$. Thus at higher relative temperatures, hole doping strongly enhances ferromagnetic fluctuations. In all three hole doped compositions the ferrimagnetic ordering occurs at high values of absolute temperature, $T/|\theta| > 0.5$ (f<2), indicating that hole doping suppresses the intrinsic magnetic frustration present in the Cr honeycomb lattice. For $T/|\theta| \leq 0.25$, the normalised plot reflects the decrease in susceptibility seen at low temperatures.

In $\beta$-$CaCr_{2-2x}Ga_{2x}O_4$ the normalized susceptibility shows a systematic change in the $T/|\theta|$ value at which deviation from Curie-Weiss behavior occurs, from $T/|\theta|\sim 0.4$ in undoped $\beta$-$CaCr_2O_4$ to $T/|\theta|\sim 1$ in $\beta$-$CaCr_{1.5}Ga_{0.5}O_4$. In all Ga doped samples, as in $\beta$-$CaCr_2O_4$, a ferromagnetic component at low $T/|\theta|$ is also observed. This transition, observed as a change in gradient and defined as $T_{co}$, occurs at higher values of $T/|\theta|$ as the concentration of Ga increases and becomes more pronounced until, for $x \geq 0.15$, the value of $C/(|\theta|\chi)$ at low $T/|\theta|$ is less than that expected for Curie-Weiss behavior. The lower temperature transition observed in the more Ga rich samples is independent of composition, and the value of $T/|\theta|\sim 0.8$ indicates that competing magnetic interactions still play a role in the magnetic transition.

It is apparent that doping $\beta$-$CaCrO_4$ has a profound effect on the magnetic interactions and the observed physical properties. In the gallium doped system, $\beta$-$CaCr_{2-2x}Ga_{2x}O_4$, antiferromagnetic interactions are still prevalent and the low temperature 3D ordering transition appears to be preserved across the entire compositional range albeit with an increasing proportion of uncompensated $Cr^{3+}$ spins. In the calcium deficient compositions hole doping results in formation of a black product with increased electrical conductivity and semiconducting behavior. This change in the electric properties indicates that the holes introduced in the calcium deficient compounds are hopping from site to site. The introduction of dynamic disorder also induces an increase in the low temperature susceptibility as a consequence of either enhanced ferromagnetic, ferrimagnetic or Dzyaloshinskii-Moriya interactions.

## 4. Conclusions

We have demonstrated that in doped $\beta$-CaCr$_2$O$_4$ compounds the type and nature of the dopant introduced plays a crucial role in the magnetic behavior observed. In both cases disruption of the Cr$^{3+}$ magnetic lattice results in increasingly ferromagnetic interactions with an increase in the magnetic susceptibility at low temperatures despite the decrease in the average magnetic moment of the cations present in the magnetic lattice. In hole doped $\beta$-Ca$_{1-y}$Cr$_2$O$_4$, the high temperature deviations appear to be ferromagnetic whereas in the $\beta$-CaCr$_{2-2x}$Ga$_{2x}$O$_4$ series it is only at lower temperatures that ferromagnetic fluctuations are observed. In the $\beta$-CaCr$_{2-2x}$Ga$_{2x}$O$_4$ series doped with non-magnetic Ga$^{3+}$ the changes in magnetic behavior occur systematically. In the calcium deficient compounds ferrimagnetic ordering is observed. From the difference in the behavior of the two related systems it is clear that hole doping has a much more dramatic effect than simply introducing disorder.

The changes in the properties of the $\beta$-CaCr$_2$O$_4$ system with the introduction of a secondary cation into the magnetic lattice, highlights the changes that can occur with only small perturbations to a low dimensional frustrated magnet. The finely tuned interactions and quasi-one-dimensional antiferromagnetic short-range interactions render the results described here specific to the $\beta$-CaCr$_2$O$_4$ family of compounds. However, since the nature of the dopant is shown to have a significant effect on the properties, the reliability in general of elucidating the behavior of magnetically frustrated systems from related disordered compositions is called into question.

**Acknowledgement**

The authors wish to thank Y.S. Hor for his assistance with electrical transport measurements. This research was supported by the US Department of Energy, Division of Basic Energy Sciences, Grant DE-FG02-08ER46544.

**Figure Captions:**

**Figure 1:** Polyhedral representation of the β-CaCr$_2$O$_4$ structure viewed along the (a) [001], (b) [010] and (c) [100] directions. The two distinct Cr octahedral positions are shown are green and blue; calcium atoms are white. The magnetic exchange interactions between the ladders, J$_a$ and J$_b$, along the ladder spines, J$_{s-1}$ and J$_{s-2}$, and between the rungs of the ladders, J$_{r-1}$ and J$_{r-2}$ are also shown.

**Figure 2:** Variation of the lattice parameters and unit cell volume as a function of Cr$^{3+}$ for β-CaCr$_{2-2x}$Ga$_{2x}$O$_4$ (red circles) and β-Ca$_{1-y}$Cr$_2$O$_4$ (black squares). Unless shown error bars are smaller than the markers.

**Figure 3:** Temperature dependence of the magnetic susceptibility per mol of Cr (upper) and its inverse (lower) for β-CaCr$_{2-2x}$Ga$_{2x}$O$_4$ in a 1 T field. Isothermal magnetization measurement for β-CaCr$_{2-2x}$Ga$_{2x}$O$_4$ measured at 2 K is inset.

**Figure 4:** Variation as a function of composition for β-CaCr$_{2-2x}$Ga$_{2x}$O$_4$ of (a) the magnetic transitions observed from magnetic susceptibility, T$_N$, and specific heat, T$_1$ and T$_2$, measurements. T$_{co}$, the temperature at which crossover from AFM to FM behavior is observed is also plotted. (b) the effective magnetic moment per Cr, μ$_{eff}$, and (c) the Weiss temperature, |θ|.

**Figure 5:** Temperature dependence of the specific heat of β-CaCr$_{2-2x}$Ga$_{2x}$O$_4$ and β-Ca$_{0.925}$Cr$_2$O$_4$.

**Figure 6:** Variation of the logarithm of the resistivity as a function of inverse temperature for β-Ca$_{0.85}$Cr$_2$O$_4$.

**Figure 7:** Temperature dependence of the magnetic susceptibility per mol of Cr for β-Ca$_{1-y}$Cr$_2$O$_4$ collected under a 0.1 T field. The inverse susceptibility is inset.

**Figure 8:** Isothermal magnetization measurement for β-Ca$_{0.925}$Cr$_2$O$_4$ measured at selected temperatures. An Arrott plot of the high temperature data is inset.

**Figure 9:** Temperature dependence of the magnetic susceptibility per mol of Cr for a calcium deficient (black squares) and gallium doped (red circles) sample of β-CaCr$_2$O$_4$ containing 92.5% Cr$^{3+}$. The inverse susceptibility is inset.

**Figure 10:** Scaled plot of the temperature dependence of the inverse susceptibility for β-$CaCr_{2-2x}Ga_{2x}O_4$ (closed symbols) and β-$Ca_{1-y}Cr_2O_4$ (open symbols). Data sets depicted in the same color contain equivalent amounts of $Cr^{3+}$.

**Table 1:** Unit Cell and Magnetic Parameters of β-CaCr$_{2-2x}$Ga$_{2x}$O$_4$.

| x | % Cr$^{3+}$ | $a$ (Å) | $b$ (Å) | $c$ (Å) | V (Å$^3$) | $\mu_{eff}$ ($\mu_B$ mol$_{Cr}^{-1}$) | θ (K) |
|---|---|---|---|---|---|---|---|
| 0.000 | 100 | 10.6151(3) | 9.0758(3) | 2.9661(1) | 285.75(2) | 4.12 | -310 |
| 0.010 | 99 | 10.6155(6) | 9.0768(6) | 2.9670(3) | 285.89(5) | 4.14 | -313 |
| 0.025 | 97.5 | 10.6126(6) | 9.0762(6) | 2.9665(2) | 285.75(5) | 4.16 | -309 |
| 0.050 | 95 | 10.6162(9) | 9.0785(9) | 2.9681(3) | 286.07(7) | 4.12 | -291 |
| 0.075 | 92.5 | 10.6071(7) | 9.0752(7) | 2.9671(2) | 285.62(5) | 4.07 | -265 |
| 0.100 | 90 | 10.6079(7) | 9.0759(7) | 2.9689(2) | 285.84(6) | 4.05 | -257 |
| 0.150 | 85 | 10.6041(5) | 9.0757(4) | 2.9681(2) | 285.65(2) | 4.04 | -232 |
| 0.200 | 80 | 10.5985(8) | 9.0759(8) | 2.9695(2) | 285.64(6) | 3.99 | -212 |
| 0.250 | 75 | 10.5982(9) | 9.0812(9) | 2.9720(3) | 286.04(7) | 3.91 | -179 |

**Table 2:** Unit Cell and Magnetic Parameters of β-Ca$_{1-y}$Cr$_2$O$_4$.

| y | % Cr$^{3+}$ | $a$ (Å) | $b$ (Å) | $c$ (Å) | V (Å$^3$) | $\mu_{eff}$ ($\mu_B$ mol$_{Cr}^{-1}$) | θ (K) |
|---|---|---|---|---|---|---|---|
| 0.000 | 100 | 10.6151(3) | 9.0758(3) | 2.9661(1) | 285.75(2) | 4.12 | -310 |
| 0.075 | 92.5 | 10.6114(8) | 9.0701(8) | 2.9568(3) | 284.58(6) | 3.46 | -110 |
| 0.100 | 90 | 10.6071(8) | 9.0678(8) | 2.9535(3) | 284.08(6) | 3.29 | -75 |
| 0.150 | 85 | 10.5990(4) | 9.0551(4) | 2.9488(3) | 283.01(3) | 3.50 | -78 |

Figure 1

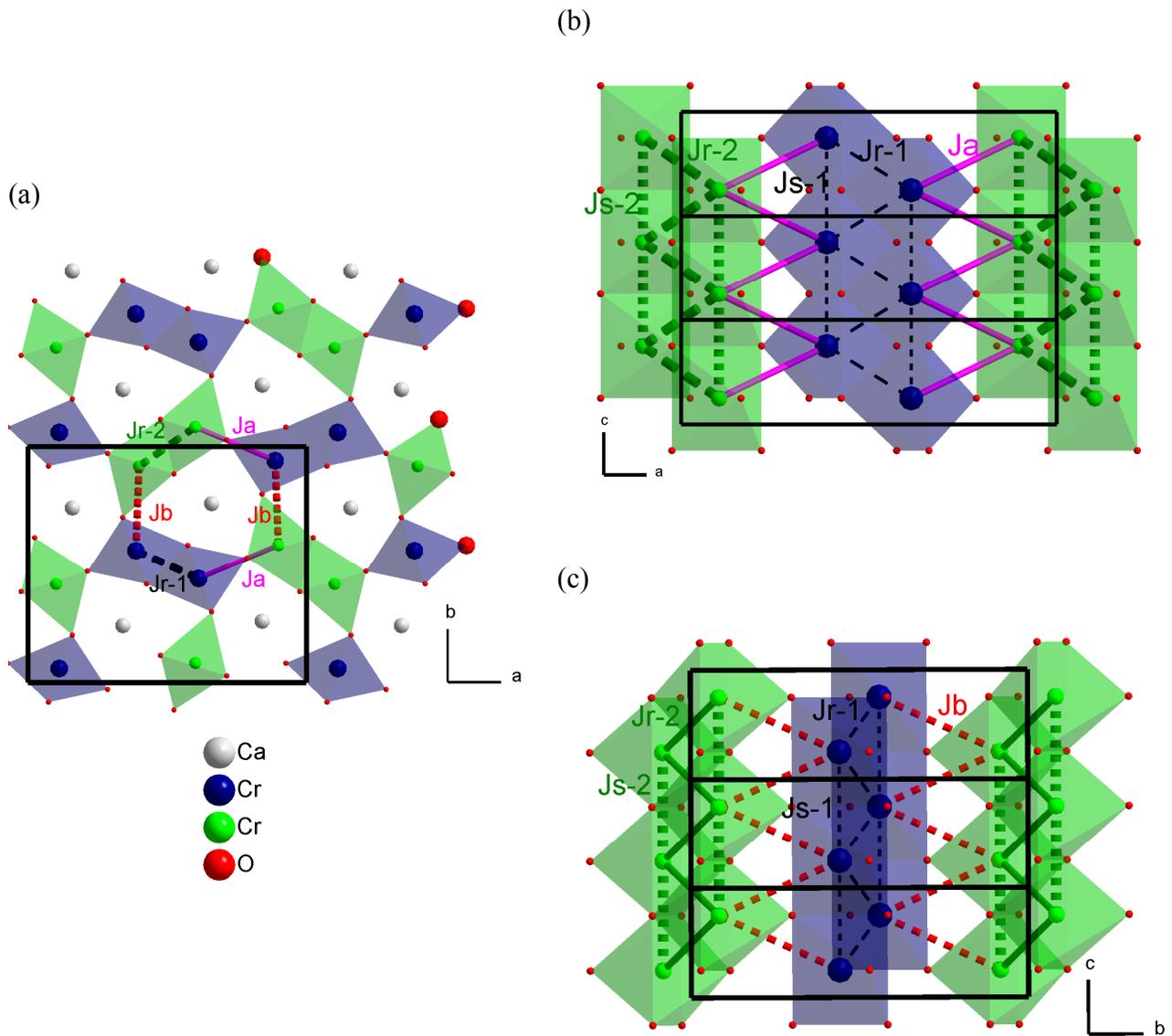

Figure 2

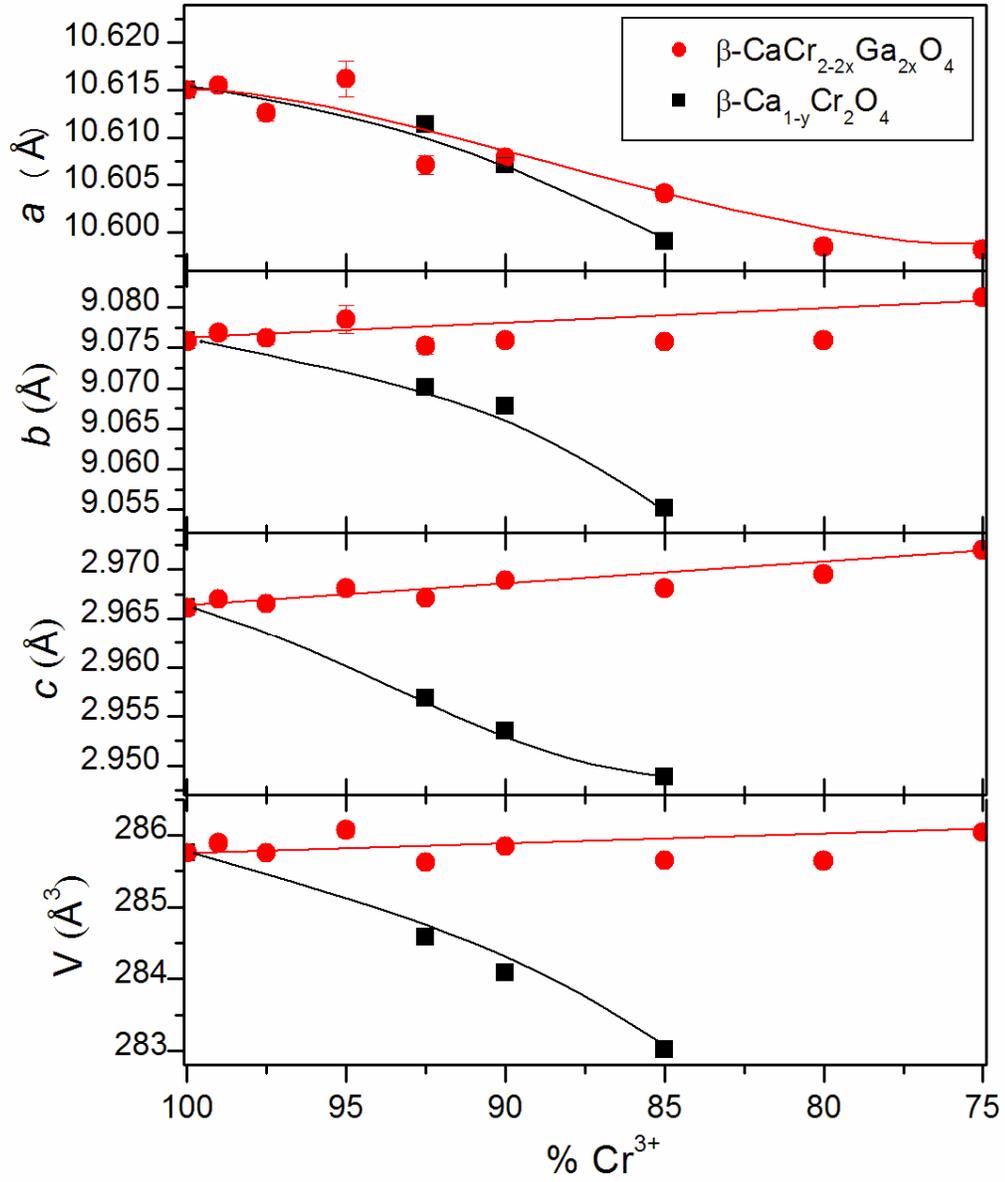

Figure 3

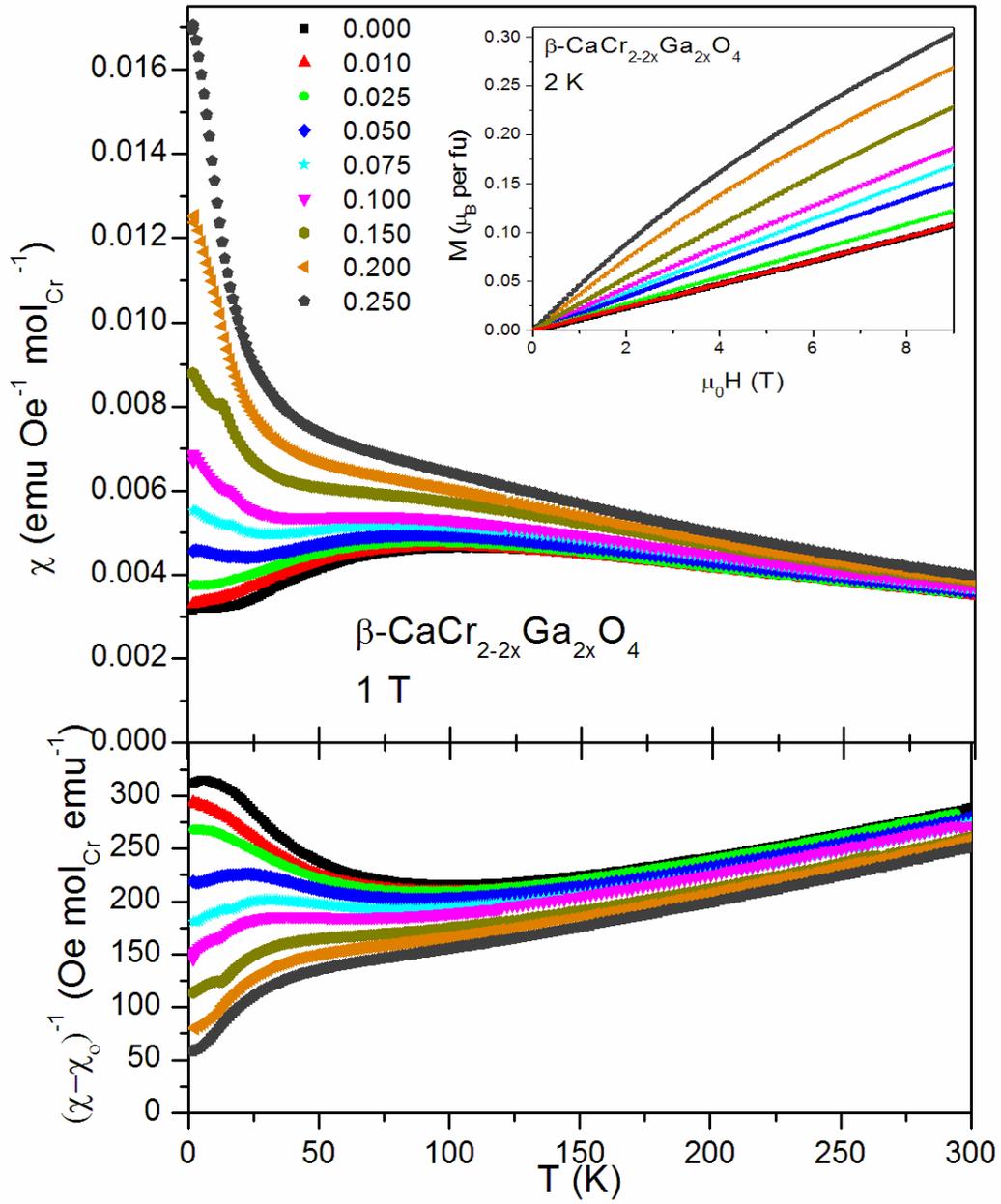

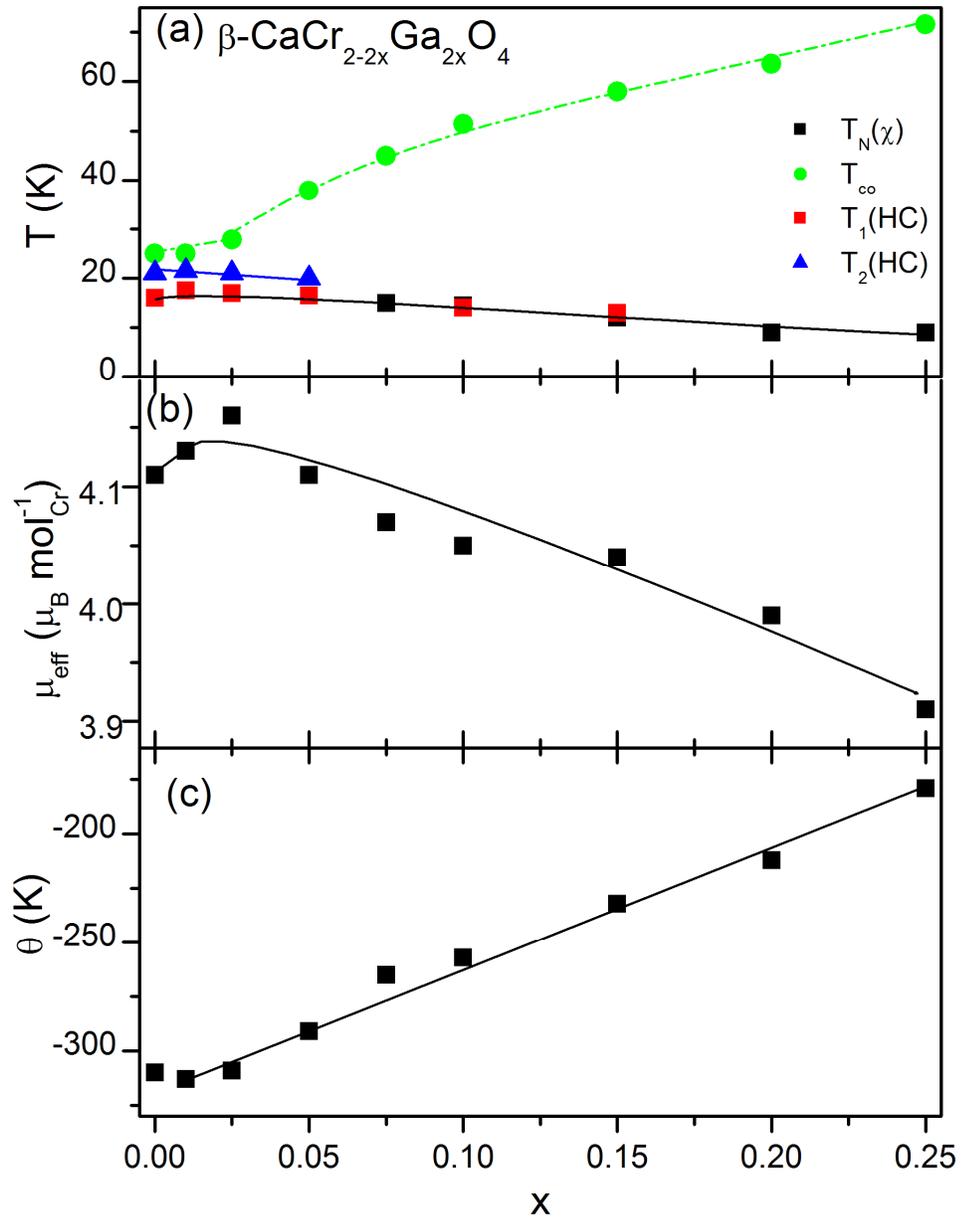

Figure 5

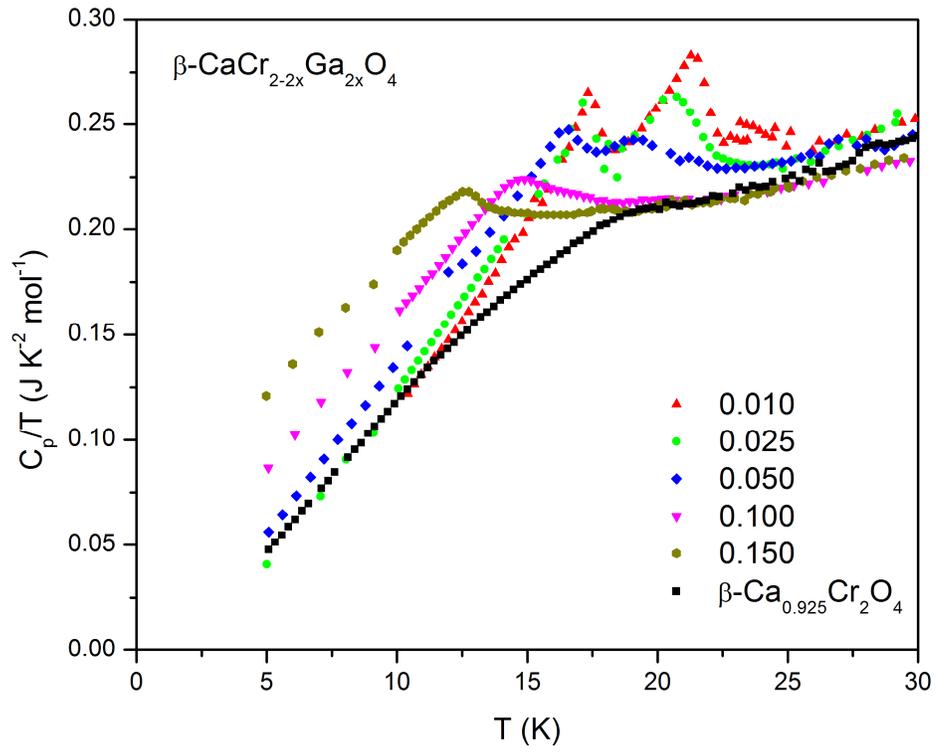

Figure 6

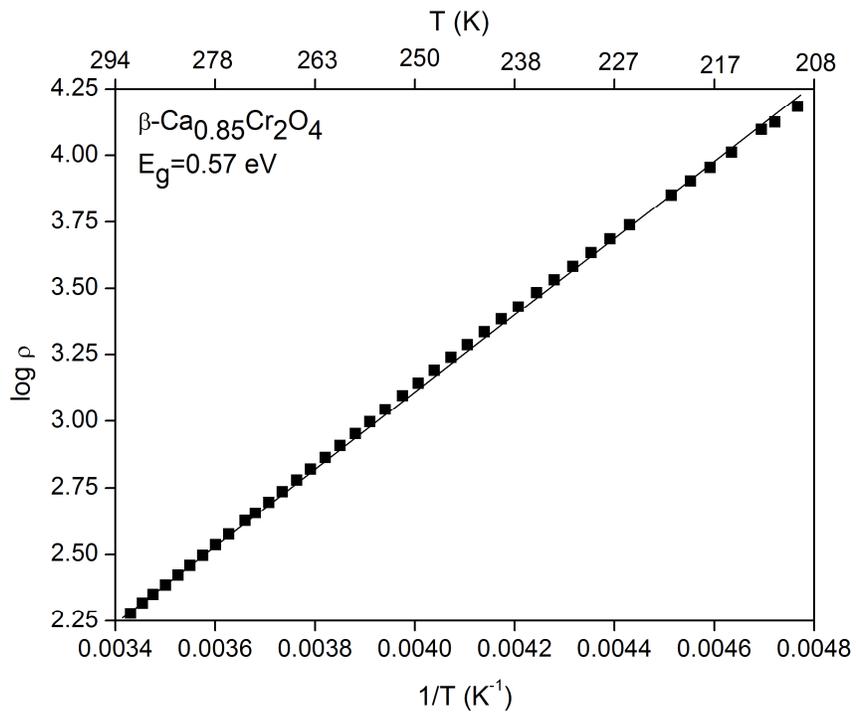

Figure 7

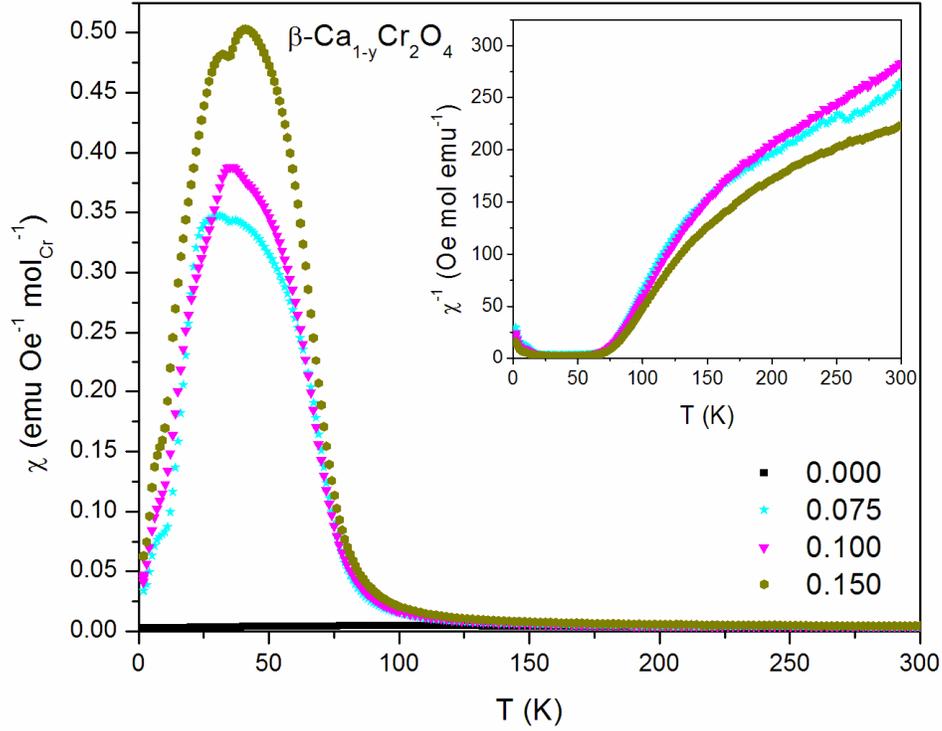

Figure 8

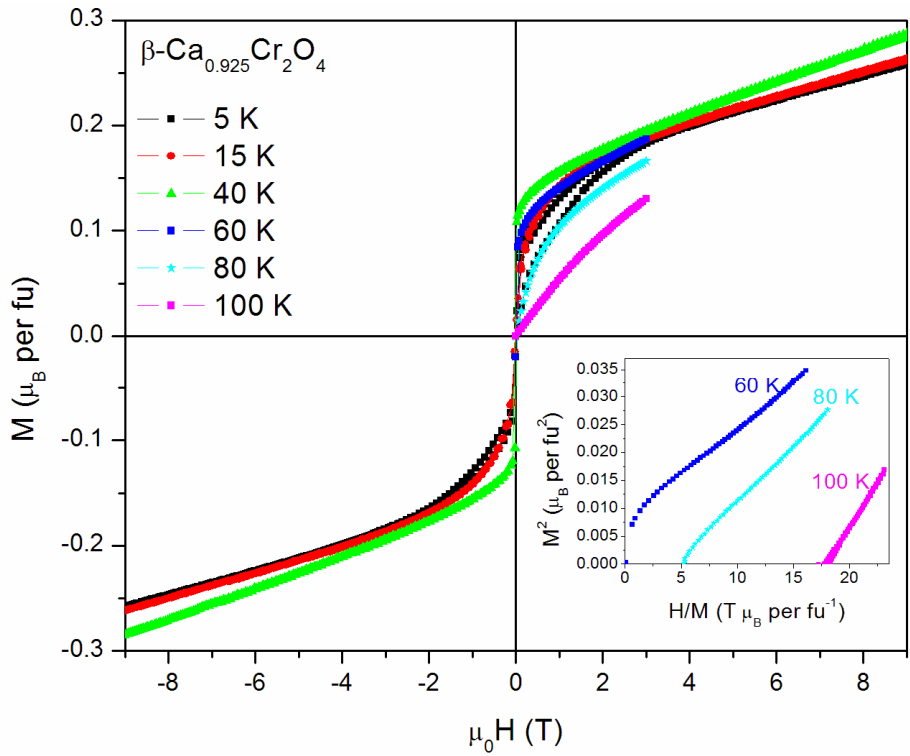

Figure 9

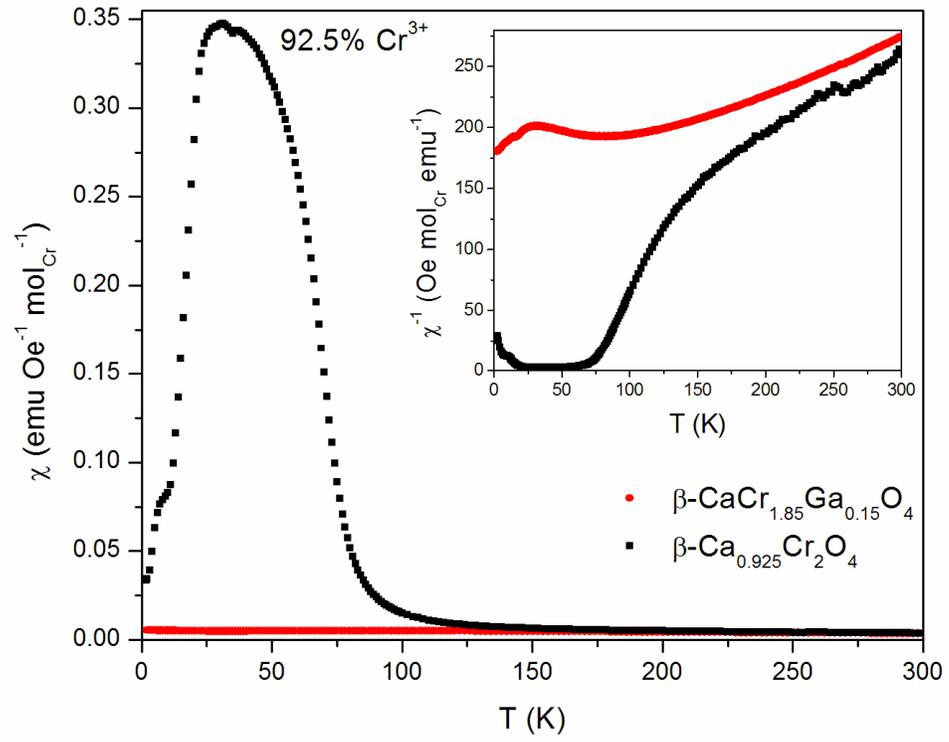

Figure 10

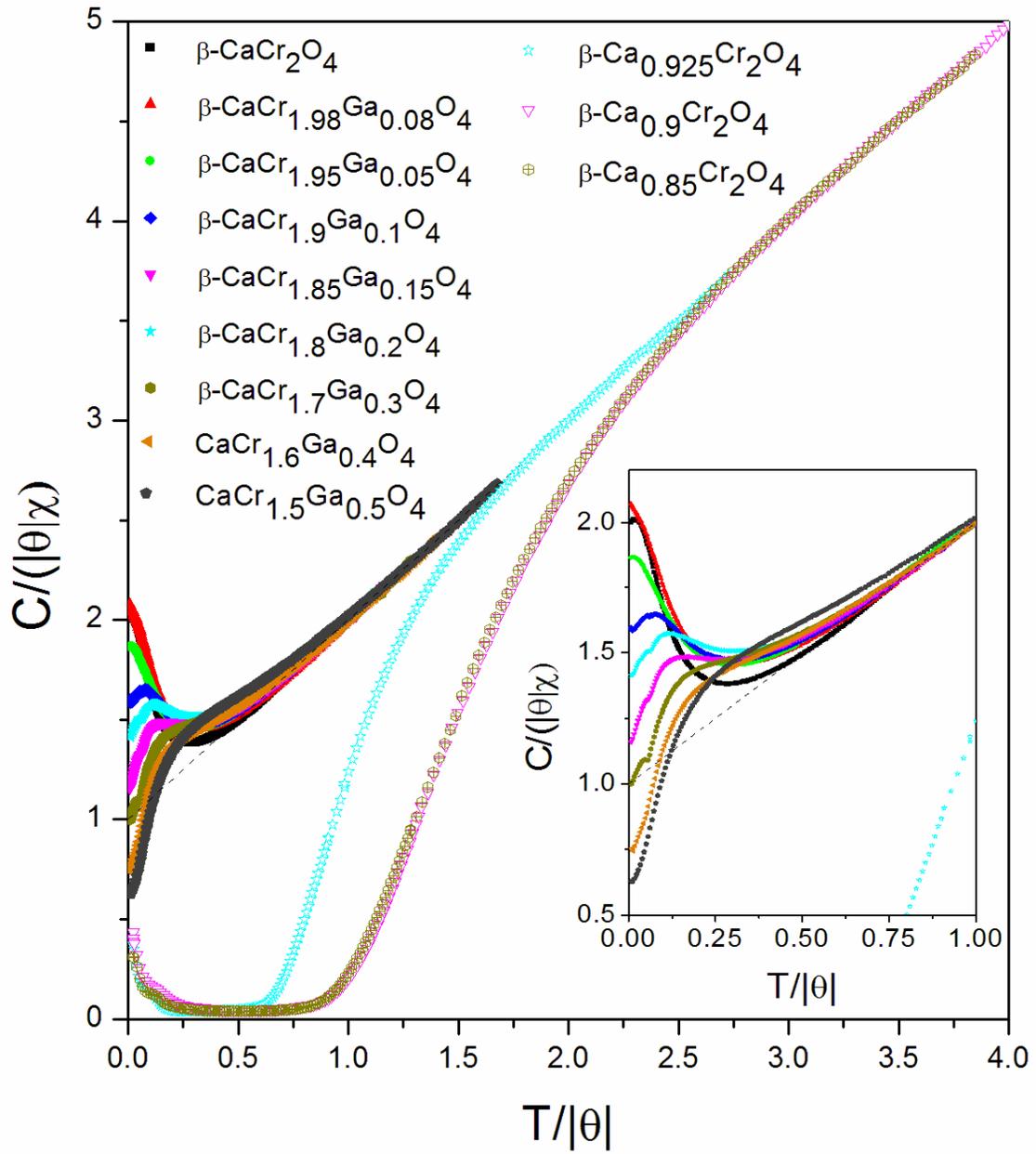